\documentclass[11pt]{article}

\usepackage{amsmath}
\usepackage{graphicx}
\usepackage{fullpage}
\usepackage{amssymb}
\usepackage{amsthm}
\usepackage{hyperref}

\newcommand{\ket}[1]{|#1\rangle}
\newcommand{\bra}[1]{\langle #1|}
\newcommand{\braket}[2]{\langle #1|#2\rangle}
\newcommand{\proj}[1]{\ket{#1} \bra{#1}}
\newcommand{\norm}[1]{\left\|#1\right\|}

\newcommand{\eps}{\epsilon}

\newcommand{\tr}{{\rm tr}\,}
\newcommand{\ot}{\otimes}
\newcommand{\bbI}{\mathbb{I}}
\newcommand{\cE}{\mathcal{E}}
\newcommand{\cF}{\mathcal{F}}

\newcommand{\cS}{\mathcal{S}}
\newcommand{\cU}{\mathcal{U}}
\newcommand{\bfm}{\mathbf{m}}
\newcommand{\bfn}{\mathbf{n}}

\newcommand{\tpiN}{\frac{2 \pi i}{N}}
\newcommand{\be}{\begin{equation}}
\newcommand{\ee}{\end{equation}}
\newcommand{\bes}{\begin{equation*}}
\newcommand{\ees}{\end{equation*}}
\def\ba#1\ea{\begin{align}#1\end{align}}
\def\bas#1\eas{\begin{align*}#1\end{align*}}
\def\bit{\begin{itemize}}
\def\eit{\end{itemize}}
\def\l{\left}
\def\r{\right}
\def\<{\langle}
\def\>{\rangle}

\newtheorem{theorem}{Theorem}
\newtheorem{lemma}[theorem]{Lemma}
\newtheorem{definition}[theorem]{Definition}
\newtheorem{corollary}[theorem]{Corollary}

\numberwithin{equation}{section}
\numberwithin{theorem}{section}

\newcommand{\nn}{\nonumber\\}
\newcommand{\eq}[1]{Eqn.~\ref{eq:#1}}

\newcommand{\thmref}[1]{Theorem \ref{thm:#1}}
\newcommand{\corref}[1]{Corollary \ref{cor:#1}}
\newcommand{\lemref}[1]{Lemma \ref{lem:#1}}
\newcommand{\secref}[1]{Section \ref{sec:#1}}
\newcommand{\defref}[1]{Definition \ref{def:#1}}

\DeclareMathOperator{\poly}{poly}
\DeclareMathOperator{\Par}{Par}
\DeclareMathOperator{\supp}{supp}
\DeclareMathOperator{\Span}{span}

\renewcommand{\iff}{\ensuremath{\mathrm{\,iff\,}}}

\def\bbC{\mathbb{C}}
\def\bbE{\mathbb{E}}

\begin{document}

\title{Efficient Quantum Tensor Product Expanders and $k$-designs}

\author{Aram W.~Harrow
\\Department of Mathematics, University of Bristol, Bristol, U.K.\\Richard A.~Low\footnote{low@cs.bris.ac.uk}\\Department of Computer Science, University of Bristol, Bristol, U.K.}

\maketitle

\begin{abstract}
Quantum expanders are a quantum analogue of expanders, and $k$-tensor product expanders are a generalisation to graphs that randomise $k$ correlated walkers.
Here we give an efficient construction of constant-degree, constant-gap
quantum $k$-tensor product expanders.  The key ingredients are an
efficient classical tensor product expander and the quantum Fourier
transform.  Our construction works whenever $k= O(n/\log n)$, where
$n$ is the number of qubits.
An immediate corollary of this result is an efficient
construction of an approximate unitary $k$-design, which is a quantum analogue of an approximate $k$-wise independent function, on $n$ qubits for any
$k=O(n/\log n)$.  Previously, no efficient constructions were known for 
$k >2$, while state designs, of which unitary designs are a generalisation,
were constructed efficiently in \cite{AmbainisEmerson07}.
\end{abstract}

\section{Introduction}

Randomness is an important resource in both classical and quantum
computing.  However, obtaining random bits is often expensive, and so
it is often desirable to minimise their use.  For example, in
classical computing, expanders and $k$-wise independent functions have
been developed for this purpose and have found wide application.  In
this paper, we explore quantum analogues of these two tools.

In quantum computing, operations are unitary gates and randomness is
often used in the form of random unitary operations.  Random unitaries
have algorithmic uses (e.g.~\cite{Sen05}) and cryptographic
applications (e.g.~\cite{AmbainisSmith04,RandomizingQuantumStates04}).
For information-theoretic applications, it is often convenient to use
unitary matrices drawn from the uniform distribution on the unitary
group (also known as the Haar measure, and described below in more
detail).  However, an $n$-qubit unitary is defined by $4^n$ real
parameters, and so cannot even be approximated implemented efficiently
using a subexponential amount of time or randomness.  Instead, we will
seek to construct efficient pseudo-random ensembles of unitaries which
resemble the Haar measure for certain applications.  For example, a
$k$-design (often referred to as a $t$-design, or a $(k,k)$-design) is
a distribution on unitaries which matches the first $k$ moments of the
Haar distribution.  This is the quantum analogue of $k$-wise
independent functions.  $k$-designs have found cryptographic uses
(e.g.~\cite{TamperResistance}) as well as physical applications
\cite{LargeDeviationskDesigns}, for which designs for large $k$ are
crucial.

Below, we will give an efficient construction of a $k$-design on $n$
qubits for any
$k$ up to $O(n/\log(n))$.  We will do this by first finding an
efficient construction of a 
quantum `$k$-copy tensor product expander' (defined later), which can
then be iterated to produce a $k$-design.  We will therefore need to
understand some of the theory of expanders before presenting our
construction.

Classical expander graphs have the property that a marker executing a
random walk on the graph will have a distribution close to the
stationary distribution after a small number of steps.  We consider a
generalisation of this, known as a $k$-tensor product expander (TPE)
and due to \cite{HastingsHarrow08}, to graphs that randomise $k$
different markers carrying out correlated random walks on the same
graph.  This is a stronger requirement than for a normal ($k=1$)
expander because the correlations between walkers (unless they start
at the same position) must be broken.  We then generalise quantum
expanders in the same way, so that the unitaries act on $k$ copies of
the system.  We give an efficient construction of a quantum $k$-TPE
which uses an efficient classical $k$-TPE as its main ingredient.  We
then give as a key application the first efficient construction of a
unitary $k$-design for any $k$.

While randomised constructions yield $k$-designs (by a modification of
Theorem 5 of \cite{TamperResistance}) and $k$-TPEs (when the dimension is
polynomially larger than $k$ \cite{HastingsHarrow08}) with near-optimal
parameters, these approaches are not efficient.  State $k$-designs,
meaning ensembles of quantum states matching the first $k$ moments of
the uniform distribution on pure states, have been efficiently
constructed in \cite{AmbainisEmerson07}, but their approach does not
appear to generalise to (unitary) $k$-designs.  Previous efficient
constructions of $k$-designs were known only for $k=1,2$, and no
efficient constant-degree, constant-gap quantum $k$-TPEs were
previously known, except for the $k=1$ case corresponding to quantum
expanders \cite{QExpandersEntropyDifference,AmbainisSmith04,AramExpanders07,QuantumMargulisExpanders}.

In \secref{exp-def}, we will define quantum expanders and other key
terms.  Then in \secref{result} we will describe our main result which
will be proved in \secref{proof}.

\subsection{Quantum Expanders}\label{sec:exp-def}

If $\cS_N$ denotes the symmetric group on $N$ objects and $\pi\in \cS_N$,
then define 
\begin{equation}
B(\pi):=\sum_{i=1}^N \ket{\pi(i)}\bra{i}
\end{equation}
to be the matrix that
permutes the basis states $\ket{1},\ldots,\ket{N}$ according to
$\pi$.

We will only consider $D$-regular expander graphs here.  We can think
of a random walk on such a graph as selecting one of $D$ permutations
of the vertices randomly at each step.  We construct the permutations
as follows.  Label the vertices from $1$ to $N$.  Then label each edge
from $1$ to $D$ so that each edge label appears exactly once on the
incoming and outgoing edges of each vertex.  This gives a set of $D$
permutations. Choosing one of these permutations at random (for some
fixed probability distribution) then defines a random walk on the graph.

We now define a classical $k$-TPE:
\begin{definition}[\cite{HastingsHarrow08}]
Let $\nu$ be a probability distribution on $\cS_N$ with support on
$\leq D$ permutations.  Then $\nu$ is an $(N, D, \lambda, k)$
classical $k$-copy tensor product expander (TPE) if
\label{def:ClassicalTPE}
\be
\left\|\bbE_{\pi\sim\nu} \l[B(\pi)^{\ot k}\r] - \bbE_{\pi\sim\cS_N}
   \l[B(\pi)^{\ot k}\r]\right\|_\infty
 = \l\| \sum_{\pi\in\cS_N} \l(\nu(\pi) - \frac{1}{N!}\r)
B(\pi)^{\ot k}\r\|_\infty
 \le \lambda.
\ee
with $\lambda < 1$.
Here $\bbE_{\pi\sim\nu}$ means the expectation over $\pi$ drawn
according to
$\nu$
and $\bbE_{\pi\sim\cS_N}$ means the expectation over $\pi$ drawn
uniformly from $\cS_N$.
\end{definition}
Here, as in the rest of the paper, the norms we use are Schatten p-norms.
Setting $k=1$ recovers the usual spectral definition of an expander.
Note that a $(N,D,\lambda,k)$ TPE is also
a $(N,D,\lambda,k')$ TPE for any $k'\leq k$.  The largest meaningful
value of $k$ is $k=N$, corresponding to the case when $\nu$ describes
a Cayley graph expander on $\cS_N$.

The degree of the map is $D=|\supp \nu|$ and the gap is $1-\lambda$.
 Ideally, the degree should be small and gap large.  To be useful,
 these should normally be independent of $N$ and possibly $k$.  We say
 that a TPE construction is efficient if it can be implemented in
 $\poly \log N$ steps.  There are known constructions of efficient
 classical TPEs.  The construction of Hoory and Brodsky
 \cite{HooryBrodsky04} provides an expander with $D = \poly \log N$
 and $\lambda = 1-1/\poly(k, \log N)$ with efficient running time.  An
 efficient TPE construction is also known, due to Kassabov
 \cite{Kassabov05}, which has constant degree and gap (independent of
 $N$ and $k$).   

Similarly, we define a quantum $k$-TPE.  First we introduce the notation
$$U^{\ot k,k} = U^{\ot k} \ot (U^*)^{\ot k}.$$ 
The distribution on the unitary group that we use is the Haar measure.  This distribution is the unique unitarily invariant distribution i.e.~the only measure $dU$ where $\int f(U) dU = \int f(UV) dU$ for all functions $f$ and unitaries $V$.  Now we define
\begin{definition}[\cite{HastingsHarrow08}]
Let $\nu$ be a distribution on $\cU(N)$, the group of $N\times N$ unitary matrices, with $D=|\supp\nu|$.  Then $\nu$ is an $(N, D, \lambda, k)$ quantum $k$-copy tensor product expander if
\label{def:QuantumTPE}
\be
\left\|\bbE_{U\sim \nu} \l[U^{\ot k,k}\r] - \bbE_{U \sim \cU(N)}
\l[U^{\ot k,k}\r] \right\|_\infty \le \lambda
\ee
with $\lambda < 1$.  Here $\bbE_{U\sim \cU(N)}$ means the expectation over $U$ drawn from the Haar measure.
\end{definition}
Again, normally we want $D$ and $\lambda$ to be constants and setting $k=1$ recovers the usual definition of a quantum expander.  Note that an equivalent statement of the above definition is that, for all $\rho$,
\be
\l\| \bbE_{U\sim\nu} \l[U^{\ot k} \rho (U^\dagger)^{\ot k}\r] - 
\bbE_{U\sim\cU(N)} \l[U^{\ot k} \rho (U^\dagger)^{\ot k}\r] 
\right\|_2 \le \lambda \norm{\rho}_2
\ee

A natural application of this is to make an efficient unitary $k$-design.  A unitary $k$-design is the same as a quantum $k$-TPE except is close in the 1-norm rather than the $\infty$-norm:
\begin{definition}
Let $\nu$ be a distribution on $\cU(N)$ with $D=|\supp\nu|$.    Say that $\nu$ is an $\eps$-approximate unitary $k$-design if
\label{def:UnitaryDesign}
\be
\left\| \bbE_{U\sim \nu} [U^{\ot k,k}]  - \bbE_{U \sim U(N)} [U^{\ot k,k}] \right\|_1 \le \eps.
\ee
\end{definition}
As for TPEs, we say that a unitary design is efficient if a $\poly \log(N)$-time algorithm exists to sample $U$ from $\nu$ and to implement $U$.

Other definitions of approximate designs are possible; for example we can use the diamond norm \cite{KSV02} between the superoperators $\hat{\cE}_{\cU(N)}^k$ and $\hat{\cE}_\nu^k$ where
\be
\hat{\cE}_{\cU(N)}^k(\rho) = \bbE_{U\sim \cU(N)} [U^{\ot k}\rho (U^\dag)^{\ot k}]
\ee
and
\be
\label{eq:EH}
\hat{\cE}_\nu^k(\rho) = \bbE_{U\sim \nu}  [U^{\ot k}\rho (U^\dag)^{\ot k}]
\ee
We can then, following \cite{HL08}, define an $\eps$-approximate $k$-design as a set of unitaries $\mathcal{U}$ with
\be
\|\hat{\cE}_{\mathcal{U}(N)}^k - \hat{\cE}_\nu^k \|_\diamond \le \eps.
\label{eq:alt-qTPE-def}\ee
While these norms are in general incomparable, our results work efficiently for both definitions and indeed for any norms that are related by a factor that is polynomial in dimension.

We can make an $\eps$-approximate unitary $k$-design from a quantum $k$-TPE with $O(k \log N)$ overhead:
\begin{theorem}
If $\mathcal{U}$ is an $(N, D, \lambda, k)$ quantum $k$-TPE then iterating the map $m=\frac{1}{\log 1/\lambda} \log \frac{N^{2k}}{\eps}$ times gives an $\eps$-approximate unitary $k$-design with $D^m$ unitaries.
\end{theorem}
\begin{proof}
Iterating the TPE $m$ times gives
\bes
\left\|\bbE_{U\sim \nu} [U^{\ot k,k}] - \bbE_{U \sim \cU(N)} 
[U^{\ot k,k}] \right\|_\infty \le \lambda^m
\ees
This implies that
\bes
\left\|\bbE_{U\sim \nu} [U^{\ot k,k}] - \bbE_{U \sim \cU(N)} 
[U^{\ot k,k}] \right\|_1 \le N^{2k} \lambda^m
\ees
We take $m$ such that $N^{2k} \lambda^m=\eps$ to give the result.
\end{proof}
We omit the analogous claim for \eq{alt-qTPE-def}, as it, and the
proof, are essentially the same.

\begin{corollary}\label{cor:k-designs}
A construction of an efficient quantum $(N, D, \lambda, k)$-TPE yields
an efficient approximate unitary $k$-design, provided $\lambda = 1 -
1/\poly \log N$.  Further, if $D$ and $\lambda$ are constants, the
number of unitaries in the design is $N^{(O(k))}$. 
\end{corollary}

Our approach to construct an efficient quantum $k$-TPE will be to take an efficient classical $2k$-TPE and mix it with a quantum Fourier transform.  The degree is thus only larger than the degree of the classical expander by one.  Since the quantum Fourier transform on $\bbC^N$ requires $\poly\log(N)$ time, it follows that if the classical expander is efficient then the quantum expander is as well.  The main technical difficulty is to show for suitable values of $k$ that the gap of the quantum TPE is not too much worse than the gap of the classical TPE.

A similar approach to ours was first used in \cite{HastingsHarrow08} to construct a quantum expander (i.e.~a 1-TPE) by mixing a classical 2-TPE with a phase.  However, regardless of the set of phases chosen, this approach will not yield quantum $k$-TPEs from classical $2k$-TPEs for any $k\geq 2$.

\subsection{Main Result} \label{sec:result}

Let $\omega=e^{2\pi i/N}$ and define the $N$-dimensional Fourier
transform to be $\cF=\frac{1}{\sqrt{N}}\sum_{m=1}^N\sum_{n=1}^N
\omega^{mn}\ket{m}\bra{n}$.  Define $\delta_\cF$ to be the
distribution on $\cU(N)$ consisting of a point mass on $\cF$.  Our
main result is that mixing $\delta_\cF$ with a classical $2k$-TPE
yields a quantum $k$-TPE for appropriately chosen $k$ and $N$.

\begin{theorem}
\label{thm:mainresult}
Let
$\nu_C$ be a classical $(N, D, 1-\eps_C, 2k)$-TPE, and for $0<p<1$, define
$\nu_Q = p \nu_C + (1-p) \delta_\cF$.  Suppose that
\be
\eps_A := 1- 2(2k)^{4k}/\sqrt{N} > 0.
\ee
Then $\nu_Q$ is a quantum $(N, D+1, 1-\eps_Q, k)$-TPE where 
\be\eps_Q \geq \frac{\eps_A}{12}\min(p\eps_C,1-p) > 0
\label{eq:eps_Q-bound}\ee
The bound in \eq{eps_Q-bound} is
 optimised when $p=1/(1+\eps_C)$, in which case we have 
\be \eps_Q \geq \frac{\eps_A\eps_C}{24}.\ee
\end{theorem}

This means that any constant-degree, constant-gap classical
$2k$-TPE gives a quantum $k$-TPE with constant degree and gap. If the
the classical TPE is efficient then the quantum TPE is as
well.  Using \corref{k-designs}, we obtain approximate unitary
$k$-designs with polynomial-size circuits.

Unfortunately the construction does not work for all dimensions; we
require that $N = \Omega((2k)^{8k})$, so that $\eps_A$ is lower-bounded by a
positive constant.  However, in applications normally $k$ is
fixed.  An interesting open problem is to find a construction that
works for all dimensions, in particular a $k=\infty$ expander.  (Most
work on  $k=\infty$ TPEs so far has focused on the $N=2$
case \cite{BG06}.) 
We suspect our construction may work for $k$ as large as $cN$ for a
small constant $c$.  On the other hand, if $2k> N$ then the gap in
our construction drops to zero.

\section{Proof of Theorem 1.6}\label{sec:proof} 
\subsection{Proof overview}
First, we introduce some notation.
Define ${\cE}_{\cS_N}^{2k} = \bbE_{\pi\sim\cS_N}[B(\pi)^{\ot 2k}]$ and 
${\cE}_{\cU(N)}^{k} = \bbE_{U\sim\cU(N)}[U^{\ot k,k}]$.  These are
both projectors onto spaces which we label $V_{\cS_N}$ and
$V_{\cU(N)}$ respectively.  Since $V_{\cU(N)}\subset V_{\cS_N}$, it
follows that ${\cE}_{\cS_N}^{2k} - {\cE}_{\cU(N)}^{k} $ is a
projector onto the space $V_0:= V_{\cS_N}\cap V_{\cU(N)}^\perp$.
We also define ${\cE}_{\nu_C}^{2k} =
\bbE_{\pi\sim \nu_C} [B(\pi)^{\ot 2k}]$ and ${\cE}_{\nu_Q}^k =
\bbE_{U\sim \nu_Q}[U^{\ot k,k}]$.

The idea of our proof is to consider ${\cE}_{\nu_C}^{2k}$ a proxy
for ${\cE}_{\cS_N}^{2k}$; if $\lambda_C$ is small enough then this
is a reasonable approximation.  Then we can restrict our attention to
vectors in $V_0$, which we would like to
show all shrink substantially under the action of our expander.  This
in turn can be reduced to showing that $\cF^{\ot k,k}$ maps any vector in
$V_0$ to a vector that has $\Omega(1)$
amplitude in $V_{S_N}^\perp$.  This last step is the most technically
involved step of the paper, and involves careful examination of the
different vectors making up $V_{\cS_N}$.

Thus, our proof reduces to two key Lemmas.  The first allows us to
substitute  ${\cE}_{\nu_C}^{2k}$ 
for ${\cE}_{\cS_N}^{2k}$ while keeping the gap constant.
\begin{lemma}[\cite{HastingsHarrow08} Lemma 1]
\label{lem:ProjUnitaryGap}
Let $\Pi$ be a projector and let $X$ and $Y$ be operators such that
$\|X\|_\infty\leq 1$, $\|Y\|_\infty\leq 1$, $\Pi X = X \Pi = \Pi$,
$\|(I-\Pi)X(I-\Pi)\|_\infty \leq 1 - \eps_C$ and 
$\|\Pi Y \Pi\|_\infty \leq 1 - \eps_A$.  Assume $0<\eps_C, \eps_A<1$.
Then for any $0<p<1$, $\|p X + (1-p)Y\|_\infty < 1$.  Specifically,
\be \| p X + (1-p)Y \|_\infty \leq 
 1-\frac{\eps_A}{12}\min(p\eps_C,1-p).
\label{eq:intermediate-norm}\ee
\end{lemma}

We will restrict to $V_{\cU(N)}^\perp$, or equivalently, subtract the
projector $\cE_{\cU(N)}^k$ from each operator.  Thus we have $X =
\cE_{\nu_C}^{2k} - \cE_{\cU(N)}^k$, $\Pi =
{\cE}_{\cS_N}^{2k} - {\cE}_{\cU(N)}^k$ and $Y=\cF^{\ot 
  k,k} - \cE_{\cU(N)}^k$.  
According to
\defref{ClassicalTPE}, we have the bound
 \be\| (I -\Pi) X (I-\Pi) \|_\infty =
 \| {\cE}_{\nu_C}^{2k} -  {\cE}_{\cS_N}^{2k}\|_\infty \le
 1-\eps_C.
\label{eq:c2k-TPE-bound}\ee
It will remain only to bound $\lambda_A := 1-\eps_A =
\|({\cE}_{\cS_N}^{2k} - 
  {\cE}_{\cU(N)}^k) \cF^{\ot k,k} ({\cE}_{\cS_N}^{2k} -
  {\cE}_{\cU(N)}^k)\|_\infty$.
\begin{lemma}
\label{lem:Gap}
For $N \ge (2k)^2$,
\be
\label{eq:Gap}
\lambda_A = \|({\cE}_{\cS_N}^{2k} -
  {\cE}_{\cU(N)}^k) \cF^{\ot k,k} ({\cE}_{\cS_N}^{2k} -
  {\cE}_{\cU(N)}^k)\|_\infty \leq 2(2k)^{4k}/\sqrt{N}.
\ee
\end{lemma}

Combining  \eq{c2k-TPE-bound}, \lemref{Gap} and
\lemref{ProjUnitaryGap} now completes the proof of
 \thmref{mainresult}.

\subsection{Action of a Classical $2k$-TPE}

We start by analysing the action of a classical $2k$-TPE.  (We
consider $2k$-TPEs rather than general $k$-TPEs since our quantum
expander construction only uses these.)  The fixed points are states which are
unchanged when acted on by $2k$ copies of any permutation matrix.
Since the same permutation is applied to all copies, any equal indices
will remain equal and any unequal indices will remain unequal.  This
allows us to identify the fixed points of the classical expander: they
are the sums over all states with the same equality and difference
constraints.  For example, for $k=1$ (corresponding to a 2-TPE), the
fixed points are $\sum_{n_1} \ket{n_1, n_1}$ and $\sum_{n_1 \ne
n_2}\ket{n_1, n_2}$ (all off-diagonal entries equal to 1).  In
general, there is a fixed point for each partition of the set $\{1, 2,
\ldots, 2k\}$ into at most $N$ non-empty parts.  If $N \ge 2k$, which
is the only case we consider, the $2k^{\text{th}}$ Bell number
$\beta_{2k}$ gives the number of such partitions (see
e.g.~\cite{EnumerativeCombinatorics}).

We now write down some more notation to further analyse this.  If
$\Pi$ is a partition of $\{1,\ldots,2k\}$, then we write $\Pi\vdash
2k$.  We will see that $\cE_{\cS_N}^{2k}$ projects onto a space
spanned by vectors labelled by partitions.  For a partition $\Pi$, say
that $(i, j) \in \Pi$ if and only if elements $i$ and $j$ are in the
same block. 
Now we can write down the fixed points of the classical expander.  Let
\be
I_\Pi = \{(n_1, \ldots, n_{2k}) : n_i = n_j \iff (i, j) \in \Pi \}.
\ee
This is a set of tuples where indices in the same block of $\Pi$ are
equal and indices in different blocks are not equal.  
The corresponding state is
\be
\ket{I_\Pi} = \frac{1}{\sqrt{|I_\Pi|}} \sum_{\bfn\in I_\Pi} \ket{\bfn}
\ee
where $\bfn = (n_1, \ldots, n_{2k})$ and $|\Pi|$ is the number of
blocks in $\Pi$.  Note that the $\{I_\Pi\}_{\Pi\vdash 2k}$ form a
partition $\{1,\ldots,N\}^{2k}$ and thus the
$\{\ket{I_\Pi}\}_{\Pi\vdash 2k}$ form an orthonormal basis for $V_{\cS_N}$.  This is because, when applying the same permutation to all indices, indices that are the same remain the same and indices that differ remain different.  This implies that \be{\cE}_{\cS_N}^{2k} = \sum_{\Pi\vdash 2k} \proj{I_\Pi}.\ee
To evaluate the normalisation, use
$| I_\Pi | = (N)_{| \Pi |}$ where $(N)_n$ is the falling factorial
$N(N-1) \ldots (N-n+1)$.  We will later find it useful to bound
$(N)_n$ with
\be \l(1-\frac{n^2}{2N}\r)N^n \leq (N)_n \leq N^n.\ee

We will also make use of the refinement partial order:
\begin{definition}
The refinement partial order $\le$ on partitions $\Pi, \Pi' \in \Par(2k, N)$ is given by
\be
\Pi \le \Pi'  \iff (i, j) \in \Pi \Rightarrow (i, j) \in \Pi'.
\ee
\end{definition}
For example, $\{\{1, 2\}, \{3\}, \{4\}\} \le \{\{1, 2, 4\}, \{3\}\}$.  Note that $\Pi \le \Pi'$ implies that $|\Pi| \ge |\Pi'|$.

\subsubsection{Turning Inequality Constraints into Equality Constraints.}

In the analysis, it will be easier to consider just equality constraints rather than both inequality and equality constraints as in $I_\Pi$.  Therefore we make analogous definitions:
\be
E_\Pi = \{(n_1, \ldots, n_{2k}) : n_i = n_j \forall (i, j) \in \Pi \}
\ee
and
\be
\ket{E_\Pi} = \frac{1}{\sqrt{|E_\Pi|}} \sum_{\bfn \in E_\Pi} \ket{\bfn}.
\ee
Then $| E_\Pi | = N^{|\Pi|}$.  For $E_\Pi$, indices in the same block are equal, as with $I_\Pi$, but indices in different blocks need not be different.

We will need relationships between $I_\Pi$ and $E_\Pi$.   First, observe that $E_\Pi$ can be written as the union of some $I_\Pi$ sets:
\be
E_\Pi = \bigcup_{\Pi' \ge \Pi} I_{\Pi'}.
\label{eq:E-I-set-rel}\ee
To see this, note that for $\bfn\in E_\Pi$, we have $n_i=n_j \forall (i,j)\in\Pi$, but we may also have an arbitrary number of additional equalities between $n_i$'s in different blocks.   The (unique) partition 
$\Pi'$ corresponding to these equalities has the property that $\Pi$ is a refinement of $\Pi'$; that is, $\Pi'\geq \Pi$.
Thus for any $\bfn\in E_\Pi$ there exists a unique $\Pi'\ge \Pi$ such
that $\bfn\in I_{\Pi'}$.  Conversely, whenever $\Pi'\geq \Pi$, we also
have $I_{\Pi'}\subseteq E_{\Pi'} \subseteq E_{\Pi}$ because each
inclusion is achieved only be relaxing constraints. 

Using \eq{E-I-set-rel}, we can obtain a useful identity involving sums
over partitions:
\be N^{|\Pi|} = |E_\Pi| = \sum_{\Pi'\geq \Pi} |I_{\Pi'}| =
\sum_{\Pi'\geq \Pi} N_{(|\Pi'|)}. \label{eq:stirling-rel}\ee
Additionally, since both sides in \eq{stirling-rel} are degree $|\Pi|$ polynomials and are equal on $\ge |\Pi|+1$ points (we can choose any $N$ in \eq{stirling-rel} with $N\geq 2k$), it
implies that $x^{|\Pi|} = \sum_{\Pi'\geq \Pi} x_{(\Pi')}$ as an identity
on formal polynomials in $x$.

The analogue of \eq{E-I-set-rel} for the states $\ket{E_\Pi}$ and $\ket{I_\Pi}$ is similar but has to account for normalisation factors.  Thus we have
\be \sqrt{|E_\Pi|} \ket{E_\Pi} =  \sum_{\Pi'\geq \Pi} \sqrt{|I_{\Pi'}|} \ket{I_{\Pi'}}.
\label{eq:E-I-state-rel}\ee

We would also like to invert this relation, and write $\ket{I_\Pi}$ as a sum over various $\ket{E_{\Pi'}}$.  Doing so will require introducing some more notation.  Define $\zeta(\Pi,\Pi')$ to be 1 if $\Pi \leq \Pi'$ and 0 if $\Pi \not\leq \Pi'$.  This can be thought of as a matrix that, with respect to the refinement ordering, has ones on the diagonal and is upper-triangular.  Thus it is also invertible.  Define $\mu(\Pi,\Pi')$ to be the matrix inverse of $\zeta$, meaning that for all $\Pi_1,\Pi_2$, we have 
$$\sum_{\Pi'\vdash 2k} \zeta(\Pi_1,\Pi') \mu(\Pi',\Pi_2) = 
\sum_{\Pi'\vdash 2k} \mu(\Pi_1,\Pi')\zeta(\Pi',\Pi_2)  = \delta_{\Pi_1,\Pi_2},$$
where $\delta_{\Pi_1,\Pi_2}=1$ if $\Pi_1=\Pi_2$ and $=0$ otherwise.
Thus, if we rewrite \eq{E-I-state-rel} as 
\be \sqrt{|E_\Pi|} \ket{E_\Pi} =  \sum_{\Pi'\vdash 2k} \zeta(\Pi,\Pi')
\sqrt{|I_{\Pi'}|} \ket{I_{\Pi'}},\ee
then we can use $\mu$ to express $\ket{I_\Pi}$ in terms of the $\ket{E_\Pi}$ as
\be \sqrt{|I_{\Pi}|} \ket{I_{\Pi}}=  \sum_{\Pi'\vdash 2k} \mu(\Pi,\Pi')
\sqrt{|E_{\Pi'}|} \ket{E_{\Pi'}}.
\label{eq:I-into-E}\ee

This approach is a generalisation of inclusion-exclusion known as
M\"obius inversion, and the function $\mu$ is called the M\"{o}bius
function (see Chapter 3 of \cite{EnumerativeCombinatorics} for more
background).  For the case of the refinement partial order, the
M\"obius function is known:
\begin{lemma}[\cite{RotaMobius}, Section 7]
\label{lem:MobiusInversionRefinement}
$$\mu(\Pi, \Pi') = (-1)^{|\Pi| - |\Pi'|} \prod_{i=1}^{| \Pi' |} (b_i-1)!$$ where $b_i$ is the number of blocks of $\Pi$ in the $i^{\text th}$ block of $\Pi'$.
\end{lemma}

We can use this to evaluate sums involving the M\"{o}bius function for the refinement order.
\begin{lemma}
\label{lem:ModMobiusSumx}
\be
\sum_{\Pi' \ge \Pi} |\mu(\Pi, \Pi')| \,  x^{|\Pi'|} = x^{(|\Pi|)}
\label{eq:MobSumClaim}\ee
where $x$ is arbitrary and $x^{(n)}$ is the rising factorial $x(x+1) \cdots (x+n-1)$.
\end{lemma}
\begin{proof}
Start with $|\mu(\Pi, \Pi')| = (-1)^{|\Pi| - |\Pi'|} \mu(\Pi, \Pi')$ to obtain
\bas
\sum_{\Pi' \ge \Pi} | \mu(\Pi, \Pi') | x^{|\Pi'|}
&=(-1)^{|\Pi|} \sum_{\Pi' \ge \Pi} \mu(\Pi, \Pi') (-x)^{|\Pi'|} 
\\  &= (-1)^{|\Pi|} \sum_{\Pi' \ge \Pi} \mu(\Pi, \Pi') 
\sum_{\Pi''\geq\Pi'} \zeta(\Pi',\Pi'') (-x)_{(|\Pi''|)}
\eas
using \eq{stirling-rel}.  Then use M\"obius inversion and $(-x)_{(n)} = (-1)^n x^{(n)}$ to prove the result.
\end{proof}

We will mostly be interested in the special case $x=1$:
\begin{corollary}
\label{cor:ModMobiusSum}
\be
\sum_{\Pi' \ge \Pi} | \mu(\Pi, \Pi') | = |\Pi|!
\ee
\end{corollary}

Using $|\mu(\Pi,\Pi')|\geq 1$ and the fact that $\Pi\geq
\{\{1\},\ldots,\{n\}\}$ for all $\Pi\vdash n$, we obtain a bound on
the total number of partitions. 
\begin{corollary}
\label{cor:BellNumberBound}
The Bell numbers $\beta_n$ satisfy $\beta_n \le n!$.
\end{corollary}

\subsection{Fixed Points of a Quantum Expander}

We now turn to $V_{\cU(N)}$, the space fixed by the quantum expander.
By Schur-Weyl duality (see
e.g.~\cite{GoodmanWallach98}), the only operators on $(\bbC^N)^{\ot
  k}$ to commute with all $U^{\ot k}$ are linear combinations of
subsystem permutations
\be S(\pi) = \sum_{n_1=1}^N\cdots \sum_{n_k=1}^N \ket{n_{\pi^{-1}(1)},\ldots
n_{\pi^{-1}(k)}}\bra{n_1,\ldots,n_k}\ee
for $\pi \in \cS_k$.  The equivalent statement for
$V_{\cU(N)}$ is that the only states invariant under all $U^{\ot k,k}$
are of the form
\be
\frac{1}{\sqrt{N^k}}
\sum_{n_1,\ldots,n_k\in[N]}\ket{n_1,\ldots,n_k,n_{\pi(1)},\ldots,n_{\pi(k)}},
\label{eq:VUN-basis}\ee
for some permutation $\pi\in\cS_k$.  Since $\cE_{\cU(N)}^k=\bbE[U^{\ot
    k,k}]$ projects onto the set of states that is invariant under all
$U^{\ot k,k}$, it follows that $V_{\cU(N)}$ is equal to the span of
    the states in \eq{VUN-basis}.

Now we relate these states to our previous notation.  
\begin{definition}
For $\pi\in\cS_k$, define the partition corresponding to $\pi$ by 
$$P(\pi) = \l\{ \{1,k+\pi(1)\}, \{2,k+\pi(2)\},\ldots,
\{k, k + \pi(k)\}\r\}.$$
\end{definition}
Then the state in \eq{VUN-basis} is simply $\ket{E_{P(\pi)}}$, and so 
\be V_{\cU(N)} = \Span\{\ket{E_{P(\pi)}} : \pi\in\cS_k\}.
\label{eq:VUN-part-basis}\ee

Note that the classical expander has many
more fixed points than just the desired $\ket{E_{P(\pi)}}$.  The main task in
constructing a quantum expander from a classical one is to modify the
classical expander to decay the fixed points that should not be fixed
by the quantum expander.

\subsection{Fourier Transform in the Matrix Element Basis}
\label{sec:FT-mat-el}

Since we make use of the Fourier transform, we will need to know how it acts on a matrix element.  We find
\bes
\cF^{\ot k,k} \ket{\bfm} = \frac{1}{N^k} \sum_\bfn \omega^{\bfm.\bfn} \ket{\bfn}
\ees
where
\be
\bfm.\bfn = m_1 n_1 + \ldots + m_kn_k - m_{k+1}n_{k+1} - \ldots
- m_{2k} n_{2k}
\ee

We will also find it convenient to estimate the matrix elements
$\bra{E_{\Pi_1}}\cF^{\ot k,k}\ket{E_{\Pi_2}}$.  The properties we require are proven in the following lemmas.

\begin{lemma}
\label{lem:EqualityConstraints}
Choose any $\Pi_1,\Pi_2 \vdash 2k$.
Let $\bfm \in \Pi_1$ and $\bfn \in \Pi_2$.
Call the free indices of $\bfm$ $\tilde{m}_i$ for $1 \le i \le |\Pi_1|$.
Then let $\bfm.\bfn = \sum_{i = 1}^{|\Pi_1|} \sum_{j=1}^{2k} \tilde{m}_i
A_{i,j} n_j$ where $A_{i,j}$ is a $|\Pi_1|\times 2k$ matrix with entries
in $\{0, 1, -1\}$ which depends on $\Pi_1$ (but not $\Pi_2$).
Then
\be \bra{E_{\Pi_1}}\cF^{\ot k,k}\ket{E_{\Pi_2}} = 
N^{-k + \frac{|\Pi_1|-|\Pi_2|}{2}}
 \sum_{\bfn \in E_{\Pi_2}} 
\mathbb{I}\left(\sum_j A_{i,j} n_j \equiv 0 \bmod{N} \, \forall \, i\right)
\label{eq:EqualityConstraints}
\ee
where $\mathbb{I}$ is the indicator function.
\end{lemma}
\begin{proof}
Simply perform the $\bfm$ sum in
\be
\label{eq:EFE}
\bra{E_{\Pi_1}}\cF^{\ot k,k}\ket{E_{\Pi_2}} = 
N^{-\l(k + \frac{|\Pi_1|+|\Pi_2|}{2}\r)}
 \sum_{\bfm \in E_{\Pi_1}} \sum_{\bfn \in E_{\Pi_2}} 
\omega^{\bfm.\bfn}
\ee
\end{proof}

\begin{lemma}
$\bra{E_{\Pi_1}}\cF^{\ot k,k}\ket{E_{\Pi_2}}$ is real and
  positive.
\end{lemma}
\begin{proof}
Since all entries in the sum in \eq{EqualityConstraints} are nonnegative and at least one
($\bfn=0$) is strictly positive, \lemref{EqualityConstraints} implies the
result.
\end{proof}

\begin{lemma}
\label{lem:E-relax}
If $\Pi_1'\leq \Pi_1$ and $\Pi_2'\leq \Pi_2$ then
\be \sqrt{|E_{\Pi_1}|\cdot|E_{\Pi_2}|}
\bra{E_{\Pi_1}}\cF^{\ot k,k}\ket{E_{\Pi_2}}
\leq \sqrt{|E_{\Pi_1'}|\cdot|E_{\Pi_2'}|}
\bra{E_{\Pi_1'}}\cF^{\ot k,k}\ket{E_{\Pi_2'}}
\label{eq:E-relax}\ee
\end{lemma}
\begin{proof}
We prove first the special case when
$\Pi_1'=\Pi_1$, but $\Pi_2'\leq \Pi_2$ is arbitrary.   Recall that
$\Pi_2'\leq \Pi_2$ implies that $E_{\Pi_2}\subseteq E_{\Pi_2'}$.  Now the
LHS of \eq{E-relax} equals
\bas N^{-k}\sum_{\bfm \in E_{\Pi_1}, \bfn \in E_{\Pi_2}} 
&\exp\left(\tpiN \bfm.\bfn\right) \\
&=  N^{|\Pi_1|-k}
 \sum_{\bfn \in E_{\Pi_2}} 
\mathbb{I}\left(\sum_j A_{i,j} n_j \equiv 0 \bmod{N} \, \forall \, i\right)
\\ & =
N^{|\Pi_1|-k}
 \sum_{\bfn \in E_{\Pi_2'}} \bbI\l(\bfn\in E_{\Pi_2}\r)
\mathbb{I}\left(\sum_j A_{i,j} n_j \equiv 0 \bmod{N} \, \forall \,
  i\right)
\\ & \leq 
N^{|\Pi_1|-k}
 \sum_{\bfn \in E_{\Pi_2'}}
\mathbb{I}\left(\sum_j A_{i,j} n_j \equiv 0 \bmod{N} \, \forall \,
  i\right)
\\ & = 
\sqrt{|E_{\Pi_1}|\,|E_{\Pi_2'}|}
\bra{E_{\Pi_1}}\cF^{\ot k,k}\ket{E_{\Pi_2'}},
\eas
as desired.
To prove \eq{E-relax} we repeat this argument, interchanging the
roles of $\Pi_1$ and $\Pi_2$ and use the fact that $\bra{E_{\Pi_1}}\cF^{\ot k,k}\ket{E_{\Pi_2}}$ is symmetric in $\Pi_1$ and $\Pi_2$.
\end{proof}
\begin{lemma}
\label{lem:E-mat-el-bound}
\item \be \bra{E_{\Pi_1}}\cF^{\ot k,k}\ket{E_{\Pi_2}} \leq 
N^{-\frac{1}{2}\l|2k - (|\Pi_1| + |\Pi_2|)\r|}
\label{eq:E-mat-el-bound}\ee
\end{lemma}
\begin{proof}
Here, there are two
cases to consider.  The simpler case is when $|\Pi_1|+|\Pi_2|\leq
2k$.  Here we simply apply the inequality
$$\sum_{\bfm \in E_{\Pi_1}, \bfn \in E_{\Pi_2}} 
\exp\left(\tpiN \bfm.\bfn\right) \leq |E_{\Pi_1}|\, |E_{\Pi_2}|
 = N^{|\Pi_1|+|\Pi_2|}$$
to \eq{EFE}, and conclude that 
$\bra{E_{\Pi_1}}\cF^{\ot k,k}\ket{E_{\Pi_2}} \leq
N^{\frac{|\Pi_1| + |\Pi_2|}{2} -k}$.

Next, we would like to prove that
\be \bra{E_{\Pi_1}}\cF^{\ot k,k}\ket{E_{\Pi_2}} \leq
N^{k-\frac{|\Pi_1| + |\Pi_2|}{2}}. 
\label{eq:desired-EFE-bound}\ee
Here we use \lemref{E-relax} with $\Pi_1'=\Pi_1$ and $\Pi_2'=\{ \{1\},
\{2\}, \ldots, \{2k\} \}$, the maximally refined partition. 
Note that $|E_{\Pi_2'}|=N^{2k}$ and $\cF^{\ot k,k}\ket{E_{\Pi_2'}} =
\ket{0}$.  Thus
\bas \bra{E_{\Pi_1}}\cF^{\ot k,k}\ket{E_{\Pi_2}}
\leq N^{k-\frac{|\Pi_2|}{2}}
\bra{E_{\Pi_1}}\cF^{\ot k,k}\ket{E_{\Pi_2'}}
=N^{k-\frac{|\Pi_2|}{2}}
\braket{E_{\Pi_1}}{0} = 
N^{k-\frac{|\Pi_1| + |\Pi_2|}{2}},
\eas
establishing \eq{desired-EFE-bound}.
\end{proof}

\begin{lemma}
\label{lem:E-not-from-perm}
\item If $\Pi_1=\Pi_2=P(\pi)$ then
$\bra{E_{\Pi_1}}\cF^{\ot k,k}\ket{E_{\Pi_2}}=1$.  If, for any $\Pi_1$, $\Pi_2$ with $|\Pi_1| + |\Pi_2| = 2k$, either condition
isn't met (i.e. either $\Pi_1\neq \Pi_2$ or there does not exist
$\pi\in\cS_k$ such that $P(\pi)=\Pi_1=\Pi_2$) then
\be \bra{E_{\Pi_1}}\cF^{\ot k,k}\ket{E_{\Pi_2}} \leq \frac{2k}{N}
\label{eq:E-not-from-perm}\ee
for $N > k$.
\end{lemma}
\begin{proof}
In \lemref{tildeA}, we introduce the $\Pi_1 \times \Pi_2$ matrix $\tilde{A}$ with the property that
\be
\bfm.\bfn = \sum_{i=1}^{|\Pi_1|}\sum_{j=1}^{|\Pi_2|}
\tilde{m}_i \tilde{A}_{i,j} \tilde{n}_j
\ee
for all $\bfm \in \Pi_1$ and $\bfn \in \Pi_2$ where $\tilde{m}_j$ and $\tilde{n}_j$ are the free indices of $\bfm$ and $\bfn$.  This is similar to the matrix $A$ introduced in \lemref{EqualityConstraints} except only the free indices of $\bfn$ are considered.

For $\Pi_1 = \Pi_2 = P(\pi)$, \lemref{tildeA} implies that $\tilde{A} = 0$, or equivalently $\bfm.\bfn = 0$ for all $\bfm, \bfn \in P(\pi)$.  Using $|\Pi_1|+|\Pi_2| = 2k$, $\bra{E_{\Pi_1}}\cF^{\ot k,k}\ket{E_{\Pi_2}} = 1$.

Otherwise we have  $(\Pi_1,\Pi_2)\not\in\{(P(\pi),P(\pi)):\pi\in\cS_k\}$ with $|\Pi_1| + |\Pi_2| = 2k$.  For all these, \lemref{tildeA} implies that $\tilde{A}$ is nonzero (for $N > k$, no entries in $\tilde{A}$ can be $>N$ or $<-N$ so $\tilde{A} \equiv 0 \bmod{N}$ is equivalent to $\tilde{A} = 0$).  Fix an $i$ for which the $i^{\text{th}}$ row of
$\tilde{A}$ is nonzero.  We wish to count the number of $(\tilde
n_1,\ldots,\tilde n_{|\Pi_2|})$ such that $\sum_j \tilde A_{i,j} \tilde
n_j \equiv 0 \bmod{N}$.  Assume that each $\tilde{A}_{i,j}$ divides $N$ and is
nonnegative; if
not, we can replace $\tilde{A}_{i,j}$ with $\text{GCD}(|\tilde{A}_{i,j}|,N)$ by a suitable
change of variable for $\tilde{n}_j$.  

Now choose an arbitrary $j$ such that $\tilde A_{i,j}\neq 0$.  For any
values of $\tilde n_1,\ldots,\tilde n_{j-1}, \\ \tilde n_{j+1},
\ldots,\tilde n_{|\Pi_2|}$, there are $|\tilde A_{i,j}| \leq  2k $
choices of $\tilde n_j$ such that $\sum_{j} \tilde{A}_{i,j} \tilde{n}_j \equiv 0
\bmod{N}$.  Thus, there are $\leq 2kN^{|\Pi_2|-1}$ choices of $\tilde n$
such that $\sum_{j} \tilde{A}_{i,j} \tilde{n}_j \equiv 0 \bmod{N}$.
Substituting this into \eq{EqualityConstraints} (which we can
trivially modify to apply for $\tilde{A}$ rather than just $A$), we
find that
$$\bra{E_{\Pi_1}}\cF^{\ot k,k}\ket{E_{\Pi_2}}
\leq \frac{2k}{N} N^{-k + \frac{|\Pi_1|+|\Pi_2|}{2}}
= \frac{2k}{N},$$
thus establishing \eq{E-not-from-perm}.
\end{proof}

\begin{lemma}
\label{lem:tildeA}
Let $\tilde{A}$ be the matrix such that
$\bfm.\bfn = \sum_{i=1}^{|\Pi_1|}\sum_{j=1}^{|\Pi_2|}
\tilde{m}_i \tilde{A}_{i,j} \tilde{n}_j$
for all $\bfm \in \Pi_1$ and $\bfn \in \Pi_2$ where $\tilde{m}_j$ and $\tilde{n}_j$ are the free indices of $\bfm$ and $\bfn$.
Then
$\tilde{A}=0$ if and only if $\Pi_1=\Pi_2 \geq P(\pi)$ for
some $\pi\in\cS_k$.
\end{lemma}
\begin{proof}
We first consider
$\Pi_1=\Pi_2=P(\pi)$ for the ``if'' direction.  Note that for any
$\bfm,\bfn\in E_{P(\pi)}$, we have
\be\bfm. \bfn = \sum_{j=1}^k m_jn_j - \sum_{j=1}^k m_{\pi(j)} n_{\pi(j)}
 = 0.
\label{eq:P-P-cancel}\ee
This implies that $\tilde{A}=0$.  Now, choose any $\Pi_1 \ge P(\pi)$ and $\Pi_2 \ge P(\pi)$.  Then for any $\bfm \in \Pi_1$ and $\bfn \in \Pi_2$, $\bfm, \bfn \in P(\pi)$.  This means \eq{P-P-cancel} holds for this case so $\tilde{A}=0$ also.

On the other hand, suppose that $\tilde{A}=0$.  We will argue that
this implies the existence of a permutation $\pi$ such that $\Pi_1, \Pi_2 \ge P(\pi)$, thus establishing the ``only
if'' direction.

Let $\Pi_{1,j}$ (resp. $\Pi_{2,j}$) denote the $j^{\text{th}}$ block
of $\Pi_1$ (resp. $\Pi_2$).  Then
$$\tilde A_{i,j} = \sum_{\substack{i'\in\Pi_{1,i}\\ j'\in \Pi_{2,j}}}
\Lambda_{i',j'},$$
where $\Lambda_{i',j'}$ is defined to be 
$$\Lambda_{i',j'} = 
\begin{cases} 1 & \text{if $i'=j'\in\{1,\ldots,k\}$}\\
-1 & \text{if $i'=j'\in\{k+1,\ldots,2k\}$}\\
0 & \text{if $i'\neq j'$}\end{cases}.
$$
If $\tilde{A}=0$ then for each $i,j$ we have
\be \l| \Pi_{1,i} \cap \Pi_{2,j} \cap \{1,\ldots,k\}\r| = 
\l| \Pi_{1,i} \cap \Pi_{2,j} \cap \{k+1,\ldots,2k\}\r|
\label{eq:meet-balanced}.\ee
Denote the {\em meet} of $\Pi_1$ and $\Pi_2$, $\Pi_1\land\Pi_2$ to be
the greatest lower bound of $\Pi_1$ and $\Pi_2$, or equivalently the
unique partition with the fewest blocks that satisfies $\Pi_1\land\Pi_2 \leq
\Pi_1$ and $\Pi_1\land\Pi_2 \leq \Pi_2$.  The blocks of 
$\Pi_1\land\Pi_2$ are simply all of the nonempty sets
$\Pi_{1,i}\cap\Pi_{2,j}$, for $i=1,\ldots,|\Pi_1|$ and
$j=1,\ldots,|\Pi_2|$.  Thus, \eq{meet-balanced} implies that each
block of $\Pi_1\land\Pi_2$ contains an equal number of indices from
$\{1,\ldots,k\}$ as it does from $\{k+1,\ldots,2k\}$.
This implies
the existence of a permutation $\pi\in\cS_k$ such that
$\{i,k+\pi(i)\}$ is contained in a single block of $\Pi_1\land\Pi_2$
for each $i=1,\ldots,k$.  Equivalently $\Pi_1\land\Pi_2\geq P(\pi)$,
implying that $\Pi_1\geq P(\pi)$ and $\Pi_2\geq P(\pi)$.
\end{proof}

\subsection{Proof of Lemma 2.2} 
\label{sec:ProofOfLemGap}

\begin{proof}

We would like to show that, for any unit vector $\ket{\psi}\in V_0$, 
$|\bra{\psi}\cF^{\ot k,k}\ket{\psi}|^2 \leq 2(2k)^{4k}/\sqrt{N}$.   Our strategy
will be to calculate the matrix elements of $\cF^{\ot k,k}$ in the
$\ket{I_\Pi}$ and $\ket{E_\pi}$ bases.  While the $\ket{I_\Pi}$ states
are orthonormal, we will see that the $\bra{E_{\Pi_1}}\cF^{\ot
  k,k}\ket{E_{\Pi_2}}$ matrix elements are easier to calculate.  We
then use  M\"obius functions to express $\ket{I_\Pi}$ in terms of
$\ket{E_\Pi}$. 

Consider the matrix $\cE_{\cS_N}^{2k} \cF^{\ot k,k}\cE_{\cS_N}^{2k}$.  It
has $k!$ unit eigenvalues, corresponding to the $k!$-dimensional space
$V_{\cU(N)}$.   Call the $k!+1^{\text{st}}$ largest eigenvalue $\lambda_A$.
We bound $\lambda_A$ with 
\ba k! + \lambda_A^2  
&\leq \tr \l(\cE_{\cS_N}^{2k} \cF^{\ot k,k}\cE_{\cS_N}^{2k}\r)^2 \nonumber
\\ & = \sum_{\Pi_1,\Pi_2\vdash 2k} 
\l|\bra{I_{\Pi_1}}\cF^{\ot k,k}\ket{I_{\Pi_2}}\r|^2
\label{eq:I-sum}.\ea

We divide the terms in \eq{I-sum} into four types.
\begin{subequations}\label{eq:I-sum-pieces}
\begin{enumerate}\item The leading-order contribution comes from the $k!$ terms
of the form $\Pi_1=\Pi_2=P(\pi)$ for $\pi\in\cS_k$.  We bound them
with the trivial upper bound
\be |\bra{I_{\Pi_1}}\cF^{\ot k,k}\ket{I_{\Pi_2}}|^2\leq 1
\ee (which turns
out to be nearly tight).
We will then show that the remaining terms are all $k^{O(k)}/N$.
\item If $|\Pi_1|+|\Pi_2| < 2k$ then
\ba \l|\bra{I_{\Pi_1}}\cF^{\ot k,k}\ket{I_{\Pi_2}}\r|^2
& = \frac{1}{|I_{\Pi_1}|\cdot |I_{\Pi_2}| N^{2k}} 
\l| \sum_{\substack{\bfm\in\Pi_1\\ \bfn\in\Pi_2}}
e^{\frac{2\pi i \bfm.\bfn}{N}} \r|^2 
\nn & \leq \frac{|I_{\Pi_1}|\cdot |I_{\Pi_2}|}{N^{2k}} 
\nn & \leq N^{|\Pi_1| + |\Pi_2| - 2k} \leq \frac{1}{N},
\ea
where in the last line we have used the fact that $|I_\Pi| \leq
|E_\Pi| = N^{|\Pi|}$.
\item If $|\Pi_1|+|\Pi_2| > 2k$ then we will show that
\be \l|\bra{I_{\Pi_1}}\cF^{\ot k,k}\ket{I_{\Pi_2}}\r|^2 
\leq \frac{4\cdot(2k!)^2}{N}
\label{eq:too-big}\ee
\item If $|\Pi_1|+|\Pi_2|=2k$ but either $\Pi_1\neq \Pi_2$ or there is
  no $\pi\in\cS_k$ satisfying $P(\pi)=\Pi_1=\Pi_2$, then we will show that
\be \l|\bra{I_{\Pi_1}}\cF^{\ot k,k}\ket{I_{\Pi_2}}\r|^2 
\leq \frac{((2k)!+2k)^2}{N^2} \leq \frac{4\cdot(2k!)^2}{N}
\label{eq:not-from-perm}\ee
\end{enumerate}
\end{subequations}
To establish these last two claims, we will find it useful to express
$\ket{I_\Pi}$ in terms of the various $\ket{E_\Pi}$ states.  

Lemmas \ref{lem:E-mat-el-bound} and \ref{lem:E-not-from-perm} can now be used together with the M\"obius
function to bound $|\bra{I_{\Pi_1}}\cF^{\ot k,k}\ket{I_{\Pi_2}}|^2$.
First, suppose $|\Pi_1|+|\Pi_2|>2k$.  Then
\ba
 \l|\bra{I_{\Pi_1}}\cF^{\ot k,k}\ket{I_{\Pi_2}}\r| & =
\l|\sum_{\substack{\Pi_1'\geq \Pi_1\\\Pi_2'\geq \Pi_2}}
\sqrt{\frac{|E_{\Pi_1'}|\, |E_{\Pi_2'}|}
{|I_{\Pi_1}|\, |I_{\Pi_2}|}}
 \mu(\Pi_1,\Pi_1')\mu(\Pi_2,\Pi_2') 
\bra{E_{\Pi_1}}\cF^{\ot k,k}\ket{E_{\Pi_2}}\r| \nonumber
\\ & \leq
\sum_{\substack{\Pi_1'\geq \Pi_1\\\Pi_2'\geq \Pi_2}}
\sqrt{\frac{|E_{\Pi_1'}|\, |E_{\Pi_2'}|}
{|I_{\Pi_1}|\, |I_{\Pi_2}|}}
\l| \mu(\Pi_1,\Pi_1')\mu(\Pi_2,\Pi_2') \r|
\bra{E_{\Pi_1'}}\cF^{\ot k,k}\ket{E_{\Pi_2'}} \nonumber
\\ & \leq
\frac{N^k}{\sqrt{|I_{\Pi_1}|\, |I_{\Pi_2}|}}
\sum_{\substack{\Pi_1'\geq \Pi_1\\\Pi_2'\geq \Pi_2}}
\l|\mu(\Pi_1',\Pi_1)\mu(\Pi_2',\Pi_2)\r| \nonumber
\ea
by \lemref{E-mat-el-bound}.  Then using by \corref{ModMobiusSum} we find
\ba
 \l|\bra{I_{\Pi_1}}\cF^{\ot k,k}\ket{I_{\Pi_2}}\r| & = \frac{N^k |\Pi_1|!\, |\Pi_2|!}{\sqrt{(N)_{|\Pi_1|}(N)_{|\Pi_2|}}}
\label{eq:penultimate-g2k}
\\ & \leq \frac{2\cdot(2k)!}{\sqrt{N}} \nonumber
\ea
In the last step, we have assumed that $4k^2<N$, so that
$(N)_{\ell}\geq N^\ell/2$ for any $\ell\leq 2k$.  We have also made
use of the fact that (still assuming $4k^2<N$) \eq{penultimate-g2k} is maximised
when $|\Pi_1| +|\Pi_2|=2k+1$, and in particular, when one of
$|\Pi_1|$, $|\Pi_2|$ is equal to $2k$ and the other is equal to 1.

A similar analysis applies to the pairs $\Pi_1,\Pi_2$ with
$|\Pi_1|+|\Pi_2|=2k$, but with
$(\Pi_1,\Pi_2)\not\in\{(P(\pi),P(\pi)):\pi\in\cS_k\}$.  In this case, 
\ba
\bra{I_{\Pi_1}}\cF^{\ot k,k}\ket{I_{\Pi_2}} 
&= \sqrt{\frac{|E_{\Pi_1}|\, |E_{\Pi_2}|}
{|I_{\Pi_1}|\, |I_{\Pi_2}|}}
\bra{E_{\Pi_1}}\cF^{\ot k,k}\ket{E_{\Pi_2}}  \,+ \nonumber \\
& 
\sum_{\Pi_1' \ge \Pi_1, \Pi'_2 \ge \Pi_2 \atop
(\Pi_1',\Pi_2')\neq (\Pi_1,\Pi_2)} 
\sqrt{\frac{|E_{\Pi_1'}|\, |E_{\Pi_2'}|}
{|I_{\Pi_1}|\, |I_{\Pi_2}|}}
\mu(\Pi_1, \Pi_1') \mu(\Pi_2, \Pi_2')
\bra{E_{\Pi_1'}}\cF^{\ot k,k}\ket{E_{\Pi_2'}}
\label{eq:strictly-greater}\ea
We now use Lemmas \ref{lem:E-not-from-perm} and \ref{lem:E-mat-el-bound} to bound each of the two terms.  For the
first term, we use \eq{E-not-from-perm} to upper bound it with
$2k/N$.  For each choice of $\Pi_1'$ and $\Pi_2'$ in the second sum,
we have $|\Pi_1'|+|\Pi_2'|\leq 2k-1$.  Thus we can upper bound
the absolute value of the second term in \eq{strictly-greater} with
\bas \frac{1}{\sqrt{|I_{\Pi_1}|\, |I_{\Pi_2}|}}
\sum_{\Pi_1' \ge \Pi_1, \Pi'_2 \ge \Pi_2 \atop
(\Pi_1',\Pi_2')\neq (\Pi_1,\Pi_2)} 
|\mu(\Pi_1, \Pi_1') \mu(\Pi_2, \Pi_2')|
N^{|\Pi_1'|+|\Pi_2'|-k}
 &\leq \frac{2\cdot |\Pi_1|!\cdot |\Pi_2|!}{N} \\ &\leq \frac{(2k)!}{N}.
\eas
We combine the two terms and square to establish \eq{not-from-perm}.

We now put together the components from \eq{I-sum-pieces} to
upper bound \eq{I-sum}, and find that
$$k! + \lambda_A^2  \leq k! + 
\beta_{2k}^2 \frac{4\cdot(2k!)^2}{N},$$
implying that $\lambda_A \leq 2\beta_{2k}(2k!)/\sqrt{N} \leq 2(2k)^{4k}/\sqrt{N}$.
  This concludes the proof of \lemref{Gap}.
\end{proof}

\section{Conclusions}

We have shown how efficient quantum tensor product expanders can be
constructed from efficient classical tensor product expanders.  This
immediately yields an efficient construction of unitary
$k$-designs for any $k$.  Unfortunately our results do not work for
all dimensions; we require the dimension $N$ to be $\Omega((2k)^{8k})$.
While tighter analysis of our construction could likely improve this,
our construction does not work for $N < 2k$.  Constructions of
expanders for all dimensions remains an open problem. 

\subsection*{Acknowledgments}
We are grateful for funding from the Army
Research Office under grant W9111NF-05-1-0294, the European Commission
under Marie Curie grants ASTQIT (FP6-022194) and QAP (IST-2005-15848),
and the U.K. Engineering and Physical Science Research Council through
``QIP IRC.''  RL would like to thank Markus Grassl and Andreas Winter
for helpful discussions.  RL is also extremely grateful to Andreas
Winter and the rest of the Centre for Quantum Technologies, National
University of Singapore, where part of this research was carried out,
for their kind hospitality.  We would also like to thank an anonymous referee for suggesting a shorter and tighter proof of \lemref{E-not-from-perm}.


\end{document}